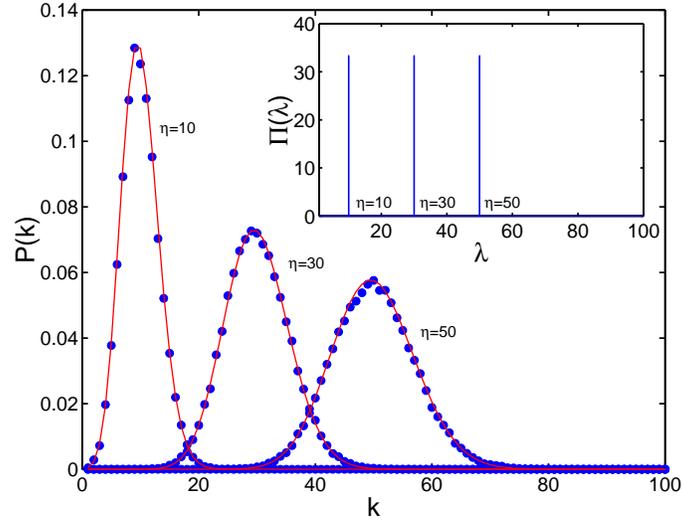

Poissonian

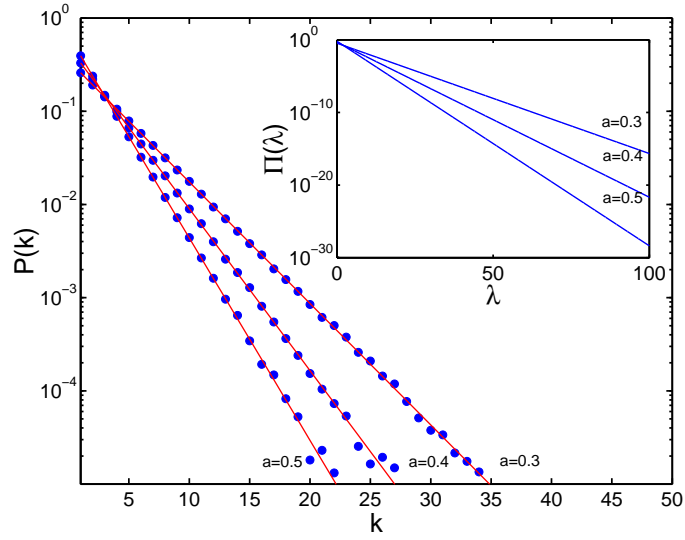

exponential

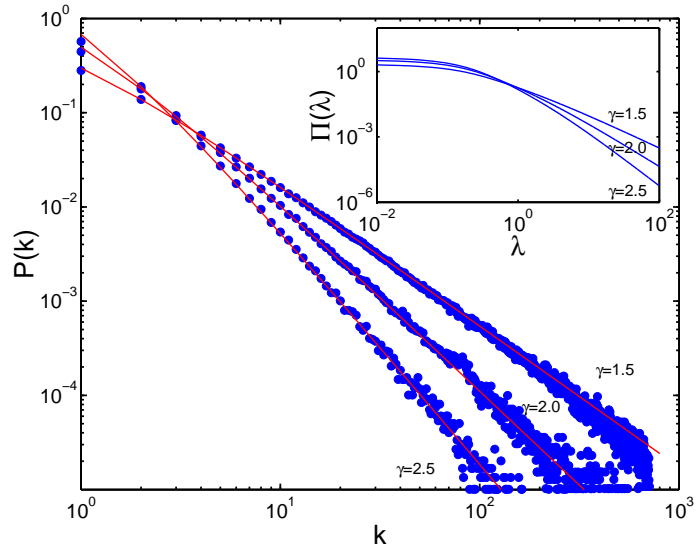

power-law

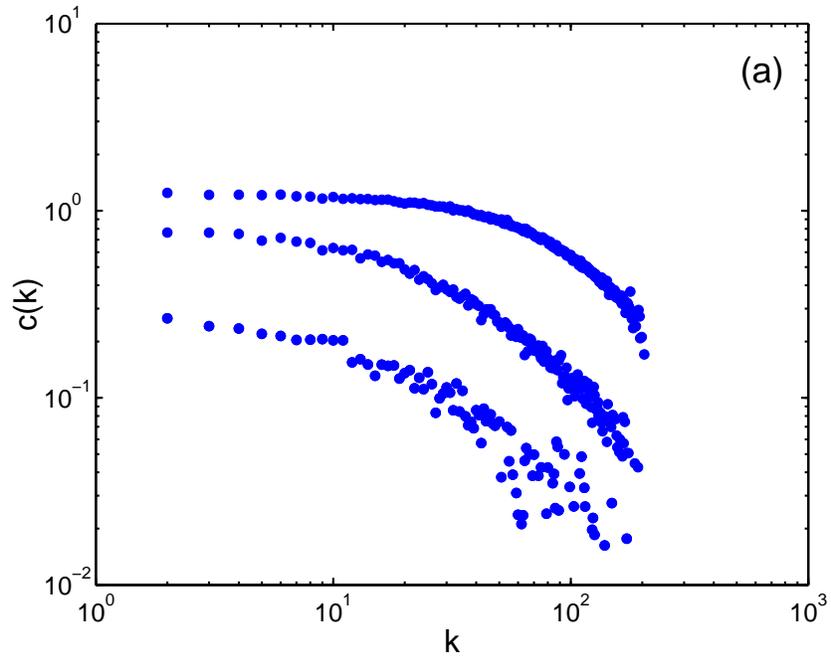

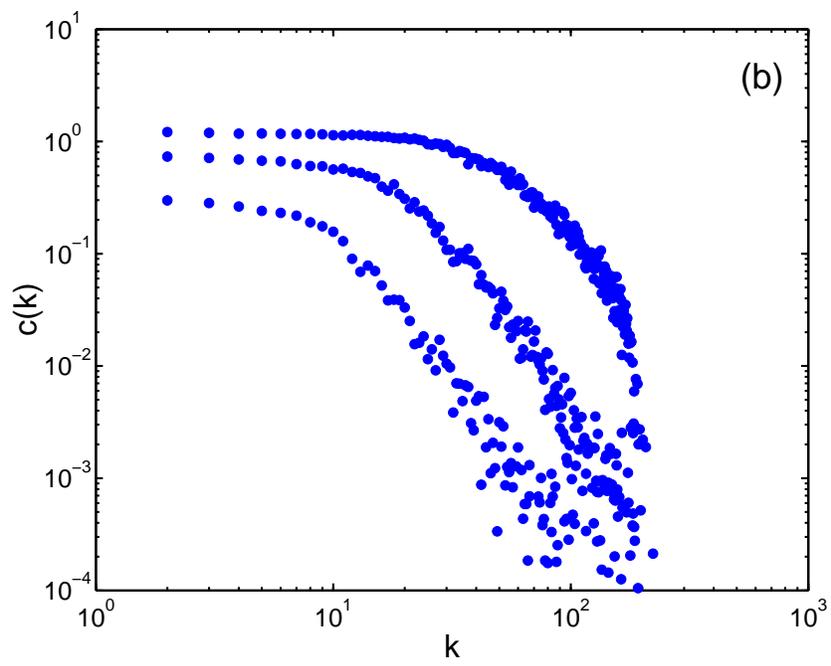

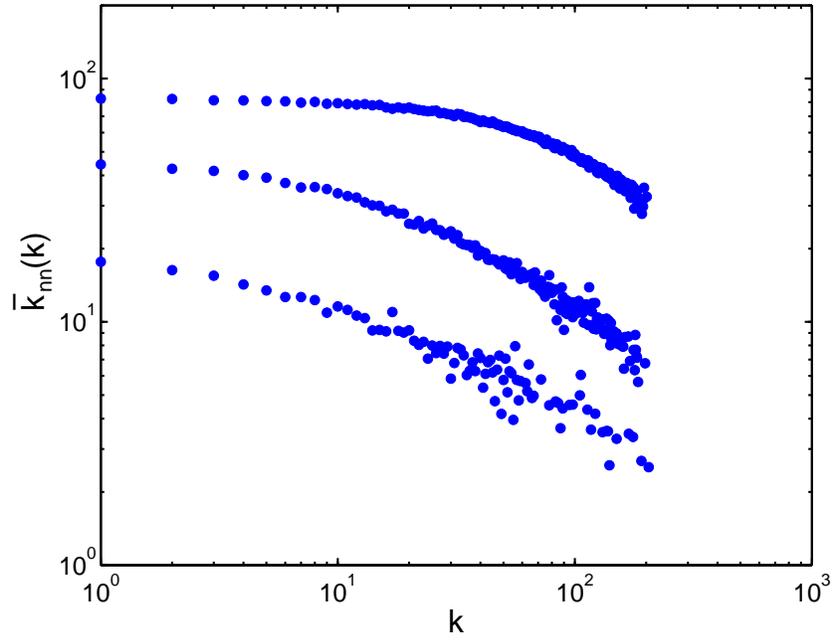

# Hierarchical and mixing properties of static complex networks emerging from the fluctuating classical random graphs

Sumiyoshi Abe[1,*] and Stefan Thurner[2,**]

[1]*Institute of Physics, University of Tsukuba, Ibaraki 305-8571, Japan*

[2]*Complex Systems Research Group, HNO, Medizinische Universität Wien*

*Währinger Gürtel 18-20, Vienna A-1090, Austria*

**Abstract**  The Erdös-Rényi classical random graph is characterized by a fixed linking probability for all pairs of vertices. Here, this concept is generalized by drawing the linking probability from a certain distribution. Such a procedure is found to lead to a static complex network with an arbitrary connectivity distribution. In particular, a scale-free network with the hierarchical organization is constructed without assuming any knowledge about the global linking structure, in contrast to the preferential attachment rule for a growing network. The hierarchical and mixing properties of the static scale-free network thus constructed are studied. The present approach establishes a bridge between a scalar characterization of individual vertices and topology of an emerging complex network. The result may offer a clue for understanding the origin of a few abundance of connectivity distributions in a wide variety of static real-world networks.

PACS number(s):    89.75.Hc, 89.75.-k, 05.70.Ln

___________________________________

*suabe@sf6.so-net.ne.jp

**thurner@univie.ac.at

In recent years, a lot of efforts have been made to characterize the structures of real-world networks observed in physical, biological, economical, and man-made systems [1-3]. An intriguing feature is that most of these networks turn out to belong to one of the following classes: small-world [4], scale-free [5], exponential, or random [6]. Note that although all of these types may involve intrinsic randomness, referred to as the "random graph" is the classical one introduced by Erdös and Rényi [6].

While the preferential attachment rule [2,5] can nicely explain how a growing network self-organizes to scale-free topology, there exist a number of real scale-free networks, which do not follow such a mechanism. Some examples are given in Ref. [7]. Of particular, interest is the static scale-free network, in which the total number of vertices does not change in time, i.e., no growth. It is therefore necessary in a coherent setting to develop a theory of static complex networks and to examine their connectivity distributions, clustering properties, and levels of hierarchy [4,8]. In such a direction, some fruitful discussions have been developed in the literature. In Refs. [9-11], a static model of varying vertex fitness has been introduced and analyzed. In analogy with the superstatistics formulation [12,13] of nonextensive statistical mechanics [14,15], a general analytic result has been presented in Ref. [16] for the relation between the linking probability distribution for a fluctuating random graph (i.e., varying fitness) and a complex network with the connectivity distribution of an arbitrary form. Another approach to static complex networks has been proposed in Refs. [17,18], in which a constant-size scale-free network is made by imposing the balance between creation and merger of vertices.

In this paper, we study the seemingly simplest modification of the Erdös-Rényi classical theory of random graphs. The classical random graph is constructed by using the linking probability which is kept fixed for all pairs of vertices. Here, we start from a statistical characterization of the linking probabilities between vertices, and arrive at static complex network with various types of connectivity, including scale-free topology. We shall not employ the ensemble picture in Ref. [16] but introduce an algorithm, which enables one to produce a single realization of each network. We are motivated by the observation of the vast abundance of similar connectivity distributions in a wide variety of static real-world networks, in which vertices may be individuals, animals, chemicals, companies and so on. Often, vertices are characterized by a linking probability defining the way of connecting them by edges. A linking probability can be nonuniform for all pairs of vertices, in general, unlike in the classical theory of random graphs. It conditions characteristics and states of vertices, and. introduces a nontrivial correlation property. We may analyze the hierarchical and mixing properties of the static complex networks thus constructed.

The present approach may offer a possibility to find a common "driving force" leading to the fact that only a few classes of networks are observed in the real world. This driving force would be featured by a linking probability distribution common in various systems. For example, many social and economic networks explicitly depend on wealth of individuals, firms, banks, and so on. Suppose that the wealth distribution, which obeys a power law common in industriarized countries [19,20], is associated with such a linking probability distribution, and assume that, other than this probability, linking is entirely random. Then, the resulting network will have a connectivity distribution, which is uniquely determined by the linking probability distribution [16]. A common driving force could thus provide an understanding of why real static networks tend to belong to only a few "universality classes".

Let us start our discussion with the classical theory of random graphs presented in the milestone papers of Erdös and Rényi [6]. There are at least two known ways to construct random graphs with $N$ vertices. The first one is to connect any two of all possible $N(N-1)/2$ pairs of vertices with probability $p \in (0,1)$ and not to connect with probability $1-p$. Then, connectivity is described by the binomial distribution, $P_{\text{rg}}(k) = \binom{N-1}{k} p^k (1-p)^{N-1-k}$, which in the large $N$ limit becomes the Poissonian

distribution

$$P_{\text{rg}}(k) = \frac{\lambda^k}{k!} e^{-\lambda}, \tag{1}$$

where $\lambda$ is the average value of connectivity, $\lambda = <k> = p(N-1)$, and is to be kept fixed. This distribution has finite moments and a characteristic scale, $\lambda$. The second one is to chose $L$ edges at random and connect their end vertices to any two vertices. In this case, $\lambda$ corresponding to that in the first construction is given by $<k> = 2L/N$. The latter construction allows the presence of tadpoles (i.e., self-interaction) and melons, in general. In the large $N$ limit with fixed $\lambda$, however, these two methods are statistically equivalent, and the resulting connectivity distributions are Poissonian as in Eq. (1).

Besides the connectivity distribution, there are several other quantities characterizing networks. One such is the clustering coefficient, $c(k)$, defined by

$$c(k) = \frac{1}{N P(k)} \sum_{i=1}^{N} c_i \, \delta_{k_i, k}, \tag{2}$$

$$c_i = \frac{2 e_i}{k_i (k_i - 1)}, \tag{3}$$

where $k_i$ is connectivity of the $i$th vertex and $P(k)$ the connectivity distribution of the network (as a simple graph) under consideration. The quantity $e_i$ is related to the total number of the edges connecting all the nearest neighbors of the $i$th vertex. Using the adjacency matrix $A = (a_{ij})$ [that is, $a_{ij} = 1$ ($a_{ij} = 0$) if the $i$th and $j$th vertices are directly connected (unconnected), and $a_{ii} = 0$ because of the absence of loops], it can be expressed as

$$e_i = (A^3)_{ii}. \tag{4}$$

For the Erdös-Rényi classical random graphs as well as the scale-free networks generated by growth with the preferential attachment rule (i.e., the Barabási-Albert networks [5]), $c(k)$ does not depend on $k$, whereas the so-called hierarchical networks [8] have $c(k)$ decaying as a power law with respect to $k$. The global clustering coefficient $C$ is defined by the average of $c_i$ over all vertices [4]. The small-world networks have the values of $C$, which are much larger than those of the Erdös-Rényi classical random graphs [4].

Another useful quantity is the correlation coefficient between connectivities of nearest-neighbor vertices [3,21]:

$$\bar{k}_{nn}(k) = \sum_{k'} k' \, P(k'|k), \tag{5}$$

If $\bar{k}_{nn}(k)$ increases (decreases) with respect to $k$, the mixing property [3,21,22] is called assortative (disassortative): that is, vertices with large values of connectivity tend to have a majority of neighboring vertices with large values (small values) of connectivity. It is known [22] that the networks of scientific coauthorships and actor collaborations have assortative mixing, whereas the Internet, the World Wide Web, protein architectures in cells, and neural networks possess disassortative mixing. Very importantly, neither the Erdös-Rényi classical random graphs nor the Barabási-Albert scale-free networks show these properties, that is, their $\bar{k}_{nn}(k)$'s are actually independent of $k$.

Once again, we notice that it is a crucial point in the Erdös-Rényi classical random graphs that each vertex has the same linking probability to get connected with one of the $N-1$ potential neighbors. Now, suppose that the linking probability is not fixed but can vary for each pair of vertices. In this case, $p$, or equivalently $\lambda$ in Eq. (1), may be a random variable obeying a certain probability distribution $\Pi(\lambda)$, which is sometimes referred to (inappropriately) as the "hidden variable distribution". Then, $P_{rg}(k)$ in Eq. (1) has to be regarded as a conditional probability distribution. Then, the correct connectivity distribution $P(k)$ is given by the marginal distribution of the joint distribution, $\Pi(\lambda) P_{rg}(k)$ [15,23]:

$$P(k) = \int_0^\infty d\lambda\, \Pi(\lambda) \frac{\lambda^k}{k!} e^{-\lambda}. \tag{6}$$

This has a nice parallelism with the superstatistics approach to nonextensive statistical mechanics [12-15]. Conventional statistical mechanics characterized by the Boltzmann factor is a theory for simple systems in chaos, whereas nonextensive statistical mechanics is designed for complex systems at the edge of chaos. Superstatistics introduces significant fluctuations of the local inverse temperature $\beta$, so that the temperature in the Boltzmann factor is stochastic and obeys a certain distribution function (commonly the chi square distribution). The resulting superposed distribution turns out to be that of nonextensive statistical mechanics. Thus, we see the following correspondence relations: (Boltzmann-Gibbs statistical mechanics)$\leftrightarrow$(random graphs), (nonextensive statistical mechanics)$\leftrightarrow$(complex networks), and (fluctuating $\beta$)$\leftrightarrow$(fluctuating $\lambda$). (For another consideration of the relation between complex networks and nonextensive statistical mechanics, see Ref. [24].)

Now, the problem is to find $\Pi(\lambda)$ for a given $P(k)$. An analytic method for solving this has recently been presented in Ref. [16]. In particular, consider a static scale-free network characterized by the power-law connectivity distribution

$$P_{sf}(k) = \frac{A}{(k + k_0)^\gamma} \tag{7}$$

with $A^{-1} = \zeta(\gamma, k_0)$, $k_0 \in (0,1)$, $\gamma > 1$, and Hurwitz's generalized zeta function $\zeta(s, a)$. Then, the exact form of $\Pi(\lambda)$ is given by [16]

$$\Pi_{sf}(\lambda) = \frac{1}{\zeta(\gamma, k_0)\Gamma(\gamma)} \int_0^\infty dt\, t^{\gamma-1} \exp\left[(1-k_0)t - (e^t - 1)\lambda\right], \qquad (8)$$

where $\Gamma(z)$ is Euler's gamma function.

Now, let us describe our algorithm for the above general construction *not in employing the ensemble picture but in aiming to construct a complex network in each realization*, in contrast to the standpoint in Ref. [16]. To be manifest, recall the following procedure of generating the Erdös-Rényi classical random graph: (i) fix the total number of vertices, $N$, (ii) fix the linking probability, $p$, for all vertices, that is, fix $\lambda = <k> = p(N-1)$ as well as the total number of edges, $L = pN(N-1)/2 = \lambda N/2$, and then, (iii) pick up the $i$th and $j$th vertices with the same probability $p$, connect them, and repeat this $L$ times. We now generalize this algorithm as follows: (I) fix the total number of vertices, $N$, (II) fix the linking probability distribution for all vertices, that is fix $\lambda_i = p_i(N-1)$ for the $i$th vertex as well as the total number of edges, $L = <p_i> N(N-1)/2 = <\lambda_i> N/2$ with the brackets being the averages with respect to the linking probability distribution [the distributions, $\Pi(p)$ and $\Pi(\lambda)$, being different from each other only by a constant scale factor of the variables, $N-1$], and then (III) pick up the $i$th and $j$th vertices with the probabilities $p_i$ and $p_j$, respectively, connect them, and repeat this $L$ times.

To realize the algorithm (I)-(III), we proceed in the following way. We assign the random variable $p_i$ drawn from the distribution $\Pi(p)$ to all $N$ vertices. Then, we impose the normalization condition, $\sum_{i=1}^N p_i = 1$, eliminating any spurious influence of the constant scale factor, $N-1$. When connecting vertices, the probability that the $i$th vertex can be connected to one of two ends of the edge is $p_i$. Here, as an example, we allow a vertex to be connected to itself twice, leading to tadpoles. This is analogous to the second method of constructing the Erdös-Rényi classical random graph with tadpoles and melons, mentioned just after Eq. (1). We have numerically ascertained that elimination of tadpoles does not affect the results reported below for the network sizes $N = 1000 \sim 10000$.

In Fig. 1, we present three different forms of $\Pi(\lambda)$: $\Pi(\lambda) = \delta(\lambda - \eta)$ with constant $\eta$, $\Pi(\lambda) = a e^{-a\lambda}$ ($a > 0$), and $\Pi(\lambda) = \Pi_{\text{sf}}(\lambda)$ in Eq. (8), which lead to the Poissonian (Erdös-Rényi classical random graph), exponential (exponential network), and power-law (scale-free network) connectivity distributions, respectively. As can be seen from Eq. (8), $\Pi_{\text{sf}}(\lambda)$ decays as a power law with the same exponent as that of the connectivity distribution in Eq. (7).

Among the above three types, most interesting may be the scale-free network. In Fig. 2, we show the plots of the clustering coefficient of the scale-free network for the three different values of the exponent: $\gamma = 1.5, 2.0, 2.5$. $c(k)$ has been calculated for each realization of the network, using the algorithm explained earlier and analyzing the corresponding adjacency matrix.

There is a technical point in calculating $c(k)$ when a given realization does not contain vertices with some values of connectivity. To treat such a case, we have examined the following two averaging methods. In the first one, averaging is carried out only for the values of connectivity present in the realization. The result obtained by this method is shown in Fig. 2 (a). The second method is that the clustering coefficient is set equal to zero for values of connectivity absent in the realization and then averaging is performed. The result based on this method is presented in Fig. 2 (b). In both cases, decay of $c(k)$ with respect to $k$ is observed, manifesting hierarchical organization of the present static scale-free networks. It is quite remarkable that $\Pi_{\text{sf}}(\lambda)$ in Eq. (8) naturally leads to such a structure, without employing any artificial treatments such as thresholding (see, e.g., Ref. [10]).

We have also ascertained that the Erdös-Rényi classical random graph corresponding to the top one in Fig. 1 has $c(k)$ that does not depend on $k$, reproducing the well known fact that that a classical random graph does not possess the hierarchical structure. However, we are not able to arrive at definite conclusions about the hierarchical structure of the exponential networks corresponding to the middle one in Fig. 1, since it turned out to be numerically very hard for our current computational power to obtain reasonable statistics for large values of connectivity, unfortunately.

Finally, in Fig. 3, we present the result about the mixing property of the static scale-free network. Corresponding to the power-law connectivity distributions in Fig. 1 (c), the correlation coefficient between connectivities of nearest-neighbor vertices, $\bar{k}_{nn}(k)$, in Eq. (5) has been calculated for different values of the exponent of the connectivity distribution: $\gamma = 1.5, 2.0, 2.5$. The decay of correlation is clearly appreciated. Consequently, we find that the static scale-free network constructed here possesses disassortative mixing.

In conclusion, we have studied generalizations of the Erdös-Rényi classical theory of random graphs to allow variable linking probabilities for pairs of vertices. In particular, we have discussed in detail the case of static scale-free networks and have seen how hierarchical organization and disassortative mixing are realized. We think that special cases of the algorithm presented here might have partially been considered in Refs. [10,11].

The present study is based on a recent work in Ref. [16], in which it is analytically shown how to construct, from the classical random graph, a static complex network with an arbitrary connectivity distribution. It is of interest to see that there exists a bridge between the global topological properties (e.g., connectivity and clustering) and the local individual ones (i.e., linking probabilities). This may potentially help to explain why diverse real-world networks have common features. Such a view is particularly attractive if the fact that the approach does not rely on global information about the entire network (such as the preferential attachment rule or its variants) is taken into account.

**ACKNOWLEDGMENTS**

S. T. acknowledges support from the Austrian Science Fund FWF P10621 G05. He also thanks the SFI and, in particular, J. D. Farmer for the hospitality extended to him.

[1]  S. N. Dorogovtsev and J. F. F. Mendes, *Evolution of Networks: From Biological Nets to the Internet and WWW* (Cambridge University Press, 2003).

[2]  R. Albert and A.-L. Barabási, Rev. Mod. Phys. **74**, 47 (2002).

[3]  R. Pastor-Satorras and A. Vespignani, *Evolution and Structure of the Internet: A Statistical Physics Approach* (Cambridge University Press, Cambridge, 2004).

[4]  D. J. Watts and S. H. Strogatz, Nature (London) **393**, 440 (1998).

[5]  A.-L. Barabási and R. Albert, Science **286**, 509 (1999).

[6]  P. Erdös and A. Rényi, Publ. Math. (Debrecen) **6**, 290 (1959); Publ. Math. Inst. Hung. Acad. Sci. **5**, 17 (1960).

[7]  S. Thurner, Europhys. News **36**, 218 (2005).

[8]  E. Ravasz and A.-L. Barabási, Phys. Rev. E **67**, 026112 (2003).

[9]  K.-I. Goh, B. Kahng, and D. Kim, Phys. Rev. Lett. **87**, 278701 (2001).

[10]  G. Caldarelli, A. Capocci, P. De Los Rios, and M. A. Muñoz, Phys. Rev. Lett. **89**, 258702 (2002).

[11]  M. Catanzaro and R. Pastor-Satorras, cond-mat/0411756.

[12]  G. Wilk and Z. Wlodarczyk, Phys. Rev. Lett. **84**, 2770 (2000).

[13]  C. Beck and E. G. D. Cohen, Physics A **322**, 267 (2003).

[14]  C. Tsallis, J. Stat. Phys. **52**, 479 (1988).

[15] *Nonextensive Statistical Mechanics and Its Applications*, edited by S. Abe and Y. Okamoto (Springer-Verlag, Heidelberg, 2001).

[16] S. Abe and S. Thurner, Phys. Rev. E **72**, 036102 (2005).

[17] B.J. Kim, A. Trusina, P. Minnhagen, and K. Sneppen, Eur. Phys. J. B **43**, 369 (2005).

[18] S. Thurner and C. Tsallis, Europhys. Lett. **72**, 197 (2005).

[19] F. Clementi and M. Gallegati, Physica A **350**, 427 (2005).

[20] W. Souma and M. Nirei, e-print physics/0505173.

[21] R. Pastor-Satorras, A. Vázquez, and A. Vespignani, Phys. Rev. Lett. **87**, 258701 (2001).

[22] M. E. J. Newman, Phys. Rev. Lett. **89**, 208701 (2002).

[23] B. Söderberg, Phys. Rev. E **66**, 066121 (2002).

[24] H. Hasegawa, e-print cond-mat/0506301, to appear in Physica A.

# Figure captions

**Fig. 1** The connectivity distributions for three different linking probabilities with $N = 10000$ vertices. The results are obtained by averaging over 10 identical network realizations. Top: the Poissonian connectivity distributions of the Erdös-Rényi classical random graphs corresponding to the Dirac delta distribution for $\Pi(\lambda)$. Middle: the exponential connectivity distributions corresponding to the exponential distributions for $\Pi(\lambda)$. Bottom: the power-law connectivity distributions of the scale-free networks corresponding to $\Pi_{sf}(\lambda)$ in Eq. (8). The solid lines represent the analytical results obtained in Ref. [16].

**Fig. 2**  The log-log plots of the dependencies of the clustering coefficient of the static scale-free networks on connectivity for three typical values of the exponent of the power-law connectivity distribution obtained by two different methods: (a) averaging is carried out only over connectivity present in each network realization, and (b) the clustering coefficient of connectivity absent in each network realization is set equal to zero and then averaging is carried out. From top to bottom in both (a) and (b): $\gamma = 1.5, 2.0, 2.5$. Averaging is performed over 100 independent realizations. Both (a) and (b) yield non-trivial hierarchical organization. Here, averaging does not mean the ensemble picture: each configuration has these behaviors (i.e., averaging is nothing but for obtaining statistically rather smooth curves).

**Fig. 3**  The log-log plots of the correlation coefficient between connectivities of nearest-neighbor vertices, $\bar{k}_{nn}(k)$, with respect to connectivity for three typical values of the exponent of the power-law connectivity distribution. From the top curve to the bottom one: $\gamma = 1.5, 2.0, 2.5$. Averaging is performed over 100 independent realizations (which are the same configurations as in Fig. 2) using the linking probability $\Pi_{sf}(\lambda)$ in Eq. (8), for networks with $N = 1000$ vertices.

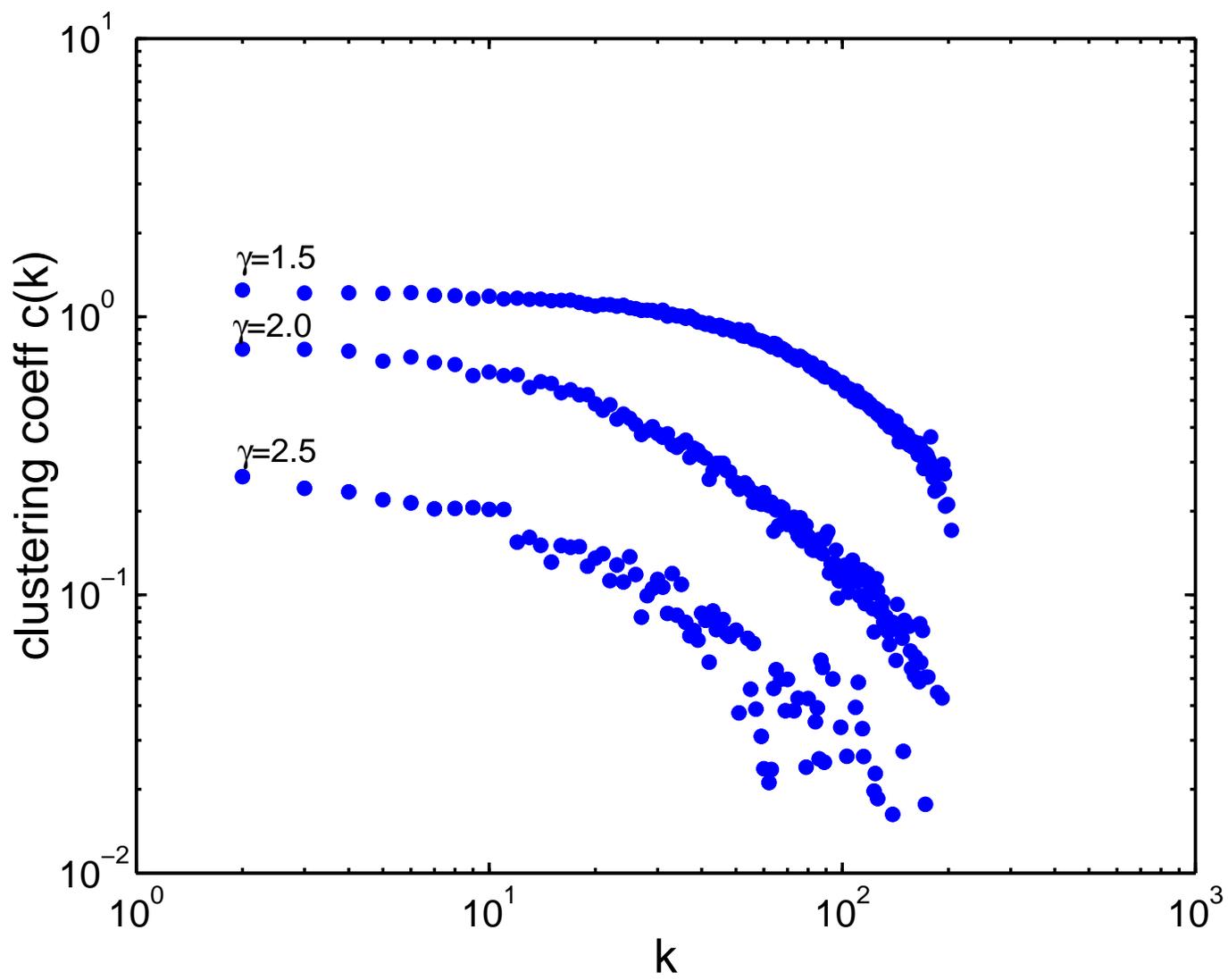